\documentclass{appolb}
\usepackage{array}
\usepackage{color}
\usepackage{graphicx}

\newcommand{\ptwo}{$\psi(2S)$}

\newcommand{\pfour}{$\psi(4S)$}

\newcommand{\uone}{$\Upsilon(1S)$}
\newcommand{\utwo}{$\Upsilon(2S)$}
\newcommand{\uthree}{$\Upsilon(3S)$}

\newcommand{\udone}{$\Upsilon(1D)$}
\newcommand{\udtwo}{$\Upsilon(2D)$}

\newcommand{\tso}{${}^{3\!}S_1$}
\newcommand{\tdo}{${}^{3\!}D_1$}
\newcommand{\tpz}{${}^{3\!}P_0$}

\newcommand{\otdo}{$1\,{}^{3\!}D_1$}
\newcommand{\ttdo}{$2\,{}^{3\!}D_1$}

\begin{document}
\title{Bottomonium vector resonances and threshold effects\thanks
{Presented by G.~Rupp at ``Excited QCD 2022'',
Giardini-Naxos, Italy, 23--29 Oct.\ 2022.}}
\author{
Eef van Beveren\footnote{Deceased on 6 December 2022.}
\address{Centro de F\'{\i}sica da UC, Departamento de
F\'{\i}sica, Universidade de Coimbra, P-3004-516, Portugal}
\\[2mm]
George Rupp\address{Centro de F\'{\i}sica e Engenharia de Materiais
Avan\c{c}ados, Instituto Superior T\'{e}cnico, Universidade de Lisboa,
P-1000-043, Portugal}
}
\maketitle
\begin{abstract}
The bottomonium spectrum is the perfect testing ground for
the confining potential and unitarisation effects.
The bottom quark is about three times heavier
than the charm quark, so that $b\bar{b}$ systems probe primarily
the short-range part of that potential. Also,
the much smaller colour-hyperfine interaction in the
$B$ mesons make the $B\bar{B}$ threshold lie significantly
higher than the $D\bar{D}$ threshold in charmonium, on a
relative scale of course.
A further complicating circumstance is that none of the
experimentally observed vector $b\bar{b}$ mesons has been
positively identified as a $^{3\!}D_1$ state, contrary to
the situation in charmonium. This makes definite conclusions
about level splittings very problematic. Finally, there are
compelling indications that the $\Upsilon(10580)$ is not
the $\Upsilon(4S)$ state, as is generally assumed. \\
Here we review an analysis of experimental bottomonium data
which show indications of the two lowest and so far unlisted
$^{3\!}D_1$ states below the $B\bar{B}$ threshold.
Next an empirical modelling of vector $b\bar{b}$ resonances
above the open-bottom threshold is revisited, 
based on the Resonance-Spectrum-Expansion production formalism
applied to other experimental data. A recent effective-Lagrangian
study supporting our non-resonant assignment of the
$\Upsilon(10580)$ is briefly discussed as well.
\end{abstract}
\section{Introduction: radial spectra of light and heavy mesons}
\label{intro}
\setcounter{figure}{0}
One of the principal goals of meson spectroscopy \cite{BR-review} is to learn
more about the confining potential, in particular its behaviour as a function
of constituent quark masses ranging from about 300--400~MeV ($u,d$) to roughly
5~GeV ($b$). Now, since this potential is generally assumed to be flavour
independent, on the basis of perturbative-QCD arguments, one would naively
expect smaller radial mass splittings for larger quark masses. However,
systems made of $u,d$ quark mostly probe the linear part of the commonly
accepted Coulomb-plus-linear (or ``funnel'') confining potential, whereas
$b\bar{b}$ states almost exclusively feel the Coulombic part. So there is
a delicate balance of two different mechanisms that will ultimately give
rise to the observed mass splittings. When the first two bottomonium states
$\Upsilon$ and $\Upsilon^\prime$ were observed, their mass splitting differing
less than 5\% from that in charmonium \cite{PDG-2022} came as quite a surprise.
While this could indeed be the accidental result of the mentioned balance,
in Ref.~\cite{Quigg-Rosner} an alternative confining potential was proposed,
namely with a logarithmic $r$ dependence. This choice trivially leads to a
radial spectrum that is independent of mass. Nevertheless, the authors 
showed \cite{Quigg-Rosner} that the funnel potential is also capable of
fitting both the first two $c\bar{c}$ and $b\bar{b}$ states, provided that
the coupling constant of the Coulombic part is strongly increased from its
fitted value in Ref.~\cite{Cornell}. In Fig.~1 the resulting $c\bar{c}$
\begin{figure}[!b]
\label{QR}
\begin{center}
\includegraphics[trim = 105mm 186mm 21mm 44mm,clip,width=9cm,angle=-0.6]
{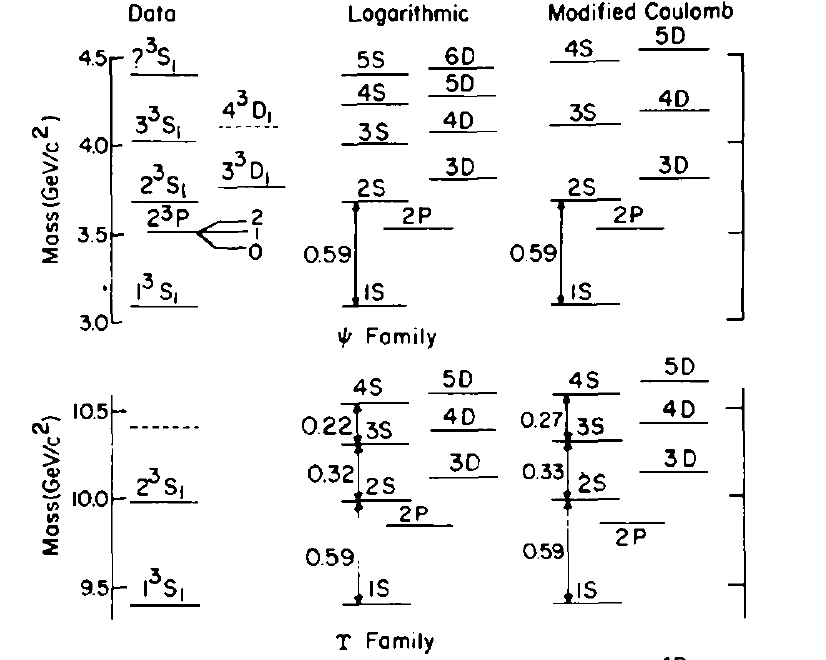} 
\end{center}
\caption{Charmonium and bottomonium spectra for Logarithmic and
``Modified Coulomb'' potentials, from Fig.~2 in Ref.~\cite{Quigg-Rosner}.
Also see text.}
\end{figure}
and $b\bar{b}$ spectra are displayed for the Logarithmic and so-called
``Modified Coulomb'' potential, together with the then \cite{Quigg-Rosner}
available data. Note that, for higher excitations, the predictions of the
two potentials clearly diverge. Moreover, the $\psi(4415)$, already observed
in 1976 \cite{PDG-2022} yet not mentioned in Ref.~\cite{Quigg-Rosner},
conflicts with the logarithmic potential, if it is indeed the \pfour\ state.

We should keep in mind that the above quarkonium potentials are very naive,
because they ignore the dynamical effects of strong decay. A clear improvement
was the coupled-channel calculation of $b\bar{b}$ states in
Ref.~\cite{Eichten}, which also used a slightly smaller coupling in the
Coulombic potential, so as to mimic asymptotic-freedom effects. Nevertheless,
this and all other funnel-type potentials will inevitably fail \cite{BR-review}
to reproduce radial spacings in the light-quark meson sector, because there
the linear part will strongly dominate and so the spacings will come out too
large. The only alternative potential that gives rise to radial splittings
independent of quark mass and also in agreement with the \ptwo, $\psi(4040)$,
and $\psi(4415)$ charmonium levels is a harmonic oscillator with universal
frequency. It was successfully applied to $c\bar{c}$ and $b\bar{b}$
vector states \cite{BR-heavy}, light, heavy-light, and heavy
vector and pseudoscalar mesons \cite{BR-light}, and light scalar mesons
\cite{BR-scalars}. In Ref.~\cite{BR-light}, the first three radial states
of the vector $\rho$, $\phi$, $\psi$, and $\Upsilon$ spectra were shown
(FIG.~1 \cite{BR-light}) to have remarkably similar mass splittings, 
especially the $\rho$, $\psi$, and $\Upsilon$ levels. Crucial for the good
model results for these mesons is a unitarised framework accounting for
non-perturbative strong-decay effects, which yield a downward mass shift of
the ground state that is larger than for the excitations, owing to the
absence of a node in the corresponding wave function \cite{BR-review}. It is
also essential that the first $\rho$ excitation be $\rho(1250)$ and not
$\rho(1450)$, as confirmed \cite{rho-1250} in a recent multichannel and
fully unitary $S$-matrix analysis with crossing-symmetry constraints.

In the present short note, we review our analyses of data published by the
BaBar Collaboration, which we believe to contain a wealth of additional
information on $b\bar{b}$ vector states. In particular, we revisit varied
evidence \cite{BR-3D1} of the so far unreported \udone\ and \udtwo\
states, as well as our analysis \cite{BR-Upsilons} of open-bottom
vector $b\bar{b}$ resonances, which suggests a non-resonant nature of the
$\Upsilon(10580)$. The latter conclusion is supported by an effective
model study \cite{Oset-Upsilons}, summarised here in conclusion.
\section{Evidence of $\Upsilon(\mbox{\ttdo})$ and indication of
$\Upsilon(\mbox{\otdo})$}
\label{QR}
In Ref.~\cite{BR-3D1} we analysed data published \cite{BABAR-3D1} 
by the BaBar Collaboration, thereby focusing on the process
$e^+e^-\to\pi^+\pi^-\Upsilon(1S)\to\pi^+\pi^-e^+e^-$. The chosen method
of analysis is to collect data on invariant $e^+e^-$ masses in bins of
10~MeV, for increasingly wide mass windows around the \uone. By looking at
the growth rate of events in each bin, fluctuations around a
smooth curve as a function of window size, and events in neighbouring bins,
clear enhancements with estimated errors and signal-to-background ratios
can be identified. For details, see Ref.~\cite{BR-3D1}.

\begin{figure}[ht]
\label{13D1-23D1}
\begin{tabular}{lr}
\hspace*{-5mm}
\includegraphics[trim = 0mm 0mm 0mm 0mm,clip,width=6.2cm,angle=0]
{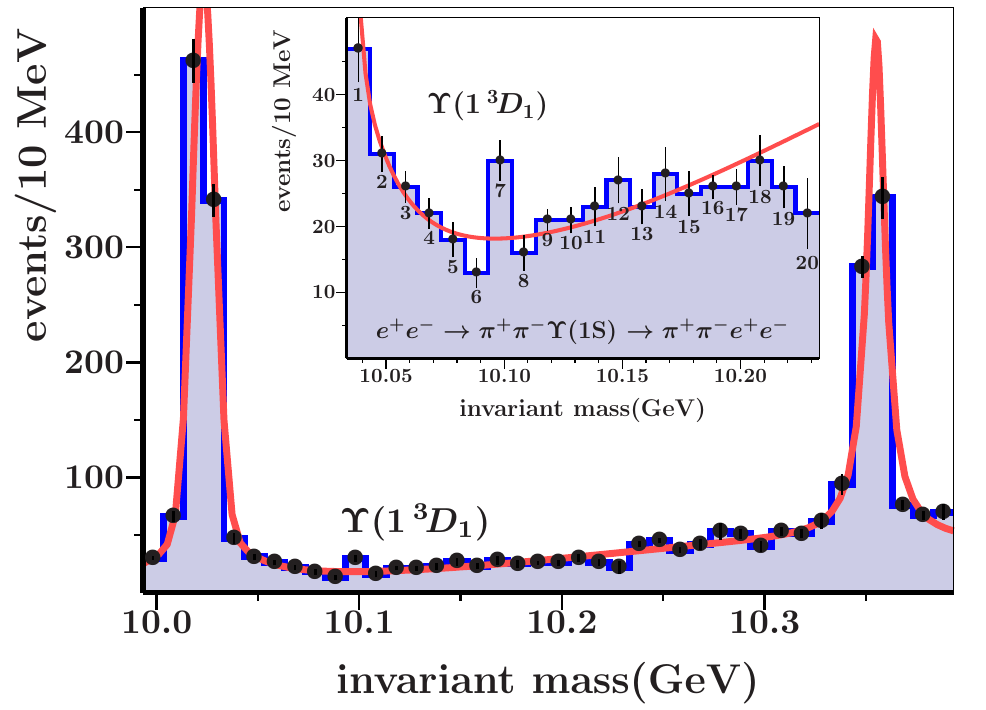} 
&
\includegraphics[trim = 0mm 0mm 0mm 0mm,clip,width=6.2cm,angle=0]
{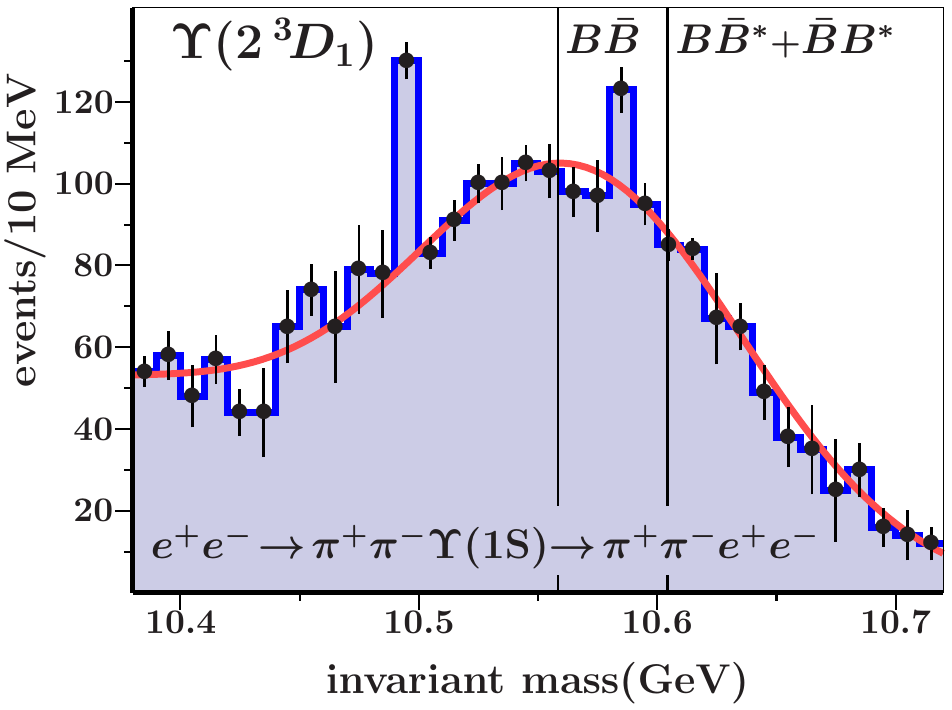}
\end{tabular}
\caption{Analysis \cite{BR-3D1} of BaBar data \cite{BABAR-3D1};
left: \udone, rigth: \udtwo. Also see text.}
\end{figure}

In Fig.~\ref{13D1-23D1} we show the graphical results of our analysis. The
left-hand plot displays a small enhancement between the two huge \utwo\ and
\uthree\ peaks, with a statistical significance of 3.0~$\sigma$, which is
in all likelihood the so far undetected \cite{PDG-2022} \udone\ state. The
7 in the enlarged inset just refers to the corresponding data bin's number.
We estimate the \udone\ mass at (10098$\,\pm\,5$)~MeV. In the right-hand plot
the energy region between roughly 10.4 and 10.7~GeV is shown, revealing a huge
peak just below 10.5~GeV, besides the known $\Upsilon(10580)$. No doubt this
amounts to the missing \cite{PDG-2022} \udtwo, whose mass we assess at
(10495$\,\pm\,5$)~MeV, with statistical significance 10.7~$\sigma$.

Notice that these inferred masses of the \udone\ and \udtwo\ states are close
to the very old coupled-channel model \cite{BR-heavy} predictions 10.14 and
10.48 GeV, respectively, and even closer \cite{BR-3D1} to the values of the
bare states at 10.113 and 10.493~GeV, respectively, as resulting from our
(with two co-authors) more general multichannel model fit in
Ref.~\cite{BR-light}. Note that mass shifts of \tdo\ states from
unitarisation are small \cite{BR-3D1} as compared to those of \tso\
states.
\section{Production analysis of excited \boldmath{$\Upsilon$} resonances}
\label{vectors}
\begin{figure}[tb]
\label{Upsilon-vectors}
\begin{center}
\includegraphics[trim = 20mm 80mm 20mm 120mm,clip,width=10.5cm,angle=0]
{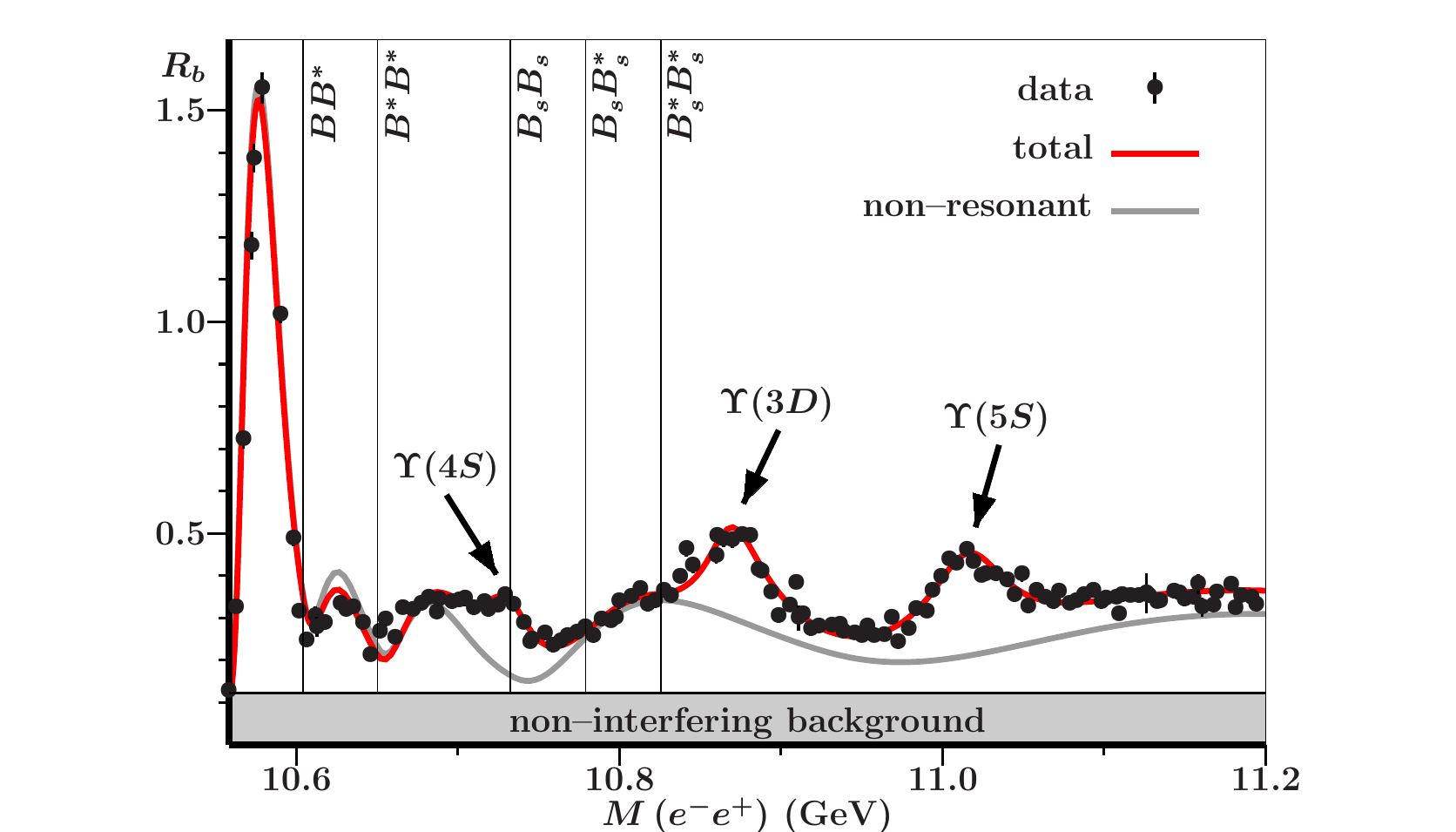} 
\end{center}
\caption{Model \cite{BR-Upsilons} fit of BaBar data \cite{BABAR-Upsilons} for
vector $b\bar{b}$ resonances. Also see text.}
\end{figure}

In Ref.~\cite{BR-Upsilons} we reanalysed another BaBar publication
\cite{BABAR-Upsilons} with a wealth of data on $b\bar{b}$ states, now above
the open-bottom threshold. In order to deal with the several partly
overlapping resonances and decay thresholds, we carried out an empirical
analysis loosely based on our multichannel production formalism
\cite{BR-production}. Its non-standard features are: non-resonant and purely
kinematical complex coefficients relating the production amplitude to the
scattering $T$-matrix while satisfying extended unitarity, with an also
kinematical, real inhomogeneous term in that relation. For the derivation of
this approach to production processes and further details, see
Ref.~\cite{BR-production}.
The results of our fit, with 16 adjustable parameters --- comparable to
a standard Breit-Wigner (BW) analysis --- are displayed in
Fig.~\ref{Upsilon-vectors} (see Ref.~\cite{BR-Upsilons} for details of the
model fit). Note that we have not included any BW parameters for the
$\Upsilon(10580)$, whose large peak is the result of the mentioned 
non-resonant lead term in our production formalism due to the opening of the
$B\bar{B}$ decay channel, further enhanced
\cite{BR-Upsilons} by the nearby subthreshold \udtwo\ pole at 10.495~GeV
\cite{BR-3D1}. From the fit we extract the true resonances
$\Upsilon(10735)$ ($\Gamma\!=\!38$~MeV), $\Upsilon(10867)$
($\Gamma\!=\!42$~MeV), and $\Upsilon(11017)$ ($\Gamma\!=\!59$~MeV), which we
interpret as $\Upsilon(4S)$, $\Upsilon(3D)$, and $\Upsilon(5S)$, respectively.
The PDG \cite{PDG-2022} lists these states with the following masses and
widths: $\Upsilon(10753)$ ($M\!=\!10753$~MeV), $\Gamma\!=\!35.5$~MeV;
$\Upsilon(10860)$ ($M\!=\!10885$~MeV), $\Gamma\!=\!37$~MeV$; \Upsilon(11020)$
($M\!=\!11000$~MeV), $\Gamma\!=\!24$~MeV.
\section{Other work on $\Upsilon$ resonances and conclusions}
In Ref.~\cite{Oset-Upsilons} a simple effective model based on the \tpz\
mechanism was employed to study $\Upsilon$ resonances above the open-bottom
threshold. From the wave-function renormalisation constant $Z$ for a propagator
dressed with loops of pairs of $B$, $B^\star$, $B_s$, and $B_s^\star$ mesons.
The authors concluded from the results that only the $\Upsilon(10580)$ has an
abnormally large meson-meson component in its wave function. Moreover, the
high peak and relatively large width of this state was argued to be
incompatible with a vector $b\bar{b}$ resonance decaying only into $B\bar{B}$
and with little phase space. These observations lend further support to our
non-resonant assignment of $\Upsilon(10580)$.

In conclusion, let us once more stress the importance for meson spectroscopy
and low-energy QCD to observe and correctly interpret the unlisted
\cite{PDG-2022} states in the bottomonium spectrum. Another interesting
approach \cite{Bicudo-Upsilons} is to extract static quark-antiquark, 
meson-meson, and transition potentials from lattice simulations and then
use these in a coupled-channel calculation. However, difficulties remain on
predicting the precise $\Upsilon(n\,$\tdo) masses. \\[5mm]

\noindent
\textit{\textbf{In Memoriam}} \\[2mm]
My longtime collaborator Eef van Beveren has
passed away due to a sudden illness on December 6th, 2022.
I will always be indebted to his brilliance in physics and
unconditional friendship.

\end{document}